\documentclass{ws-ijgmmp}
\usepackage[utf8]{inputenc}
\usepackage[T1]{fontenc}
\usepackage[colorlinks,hyperindex,plainpages=false]{hyperref}
\usepackage{amsmath}
\usepackage{amssymb}
\usepackage{amsfonts}

\newcommand{\func}[5]{\begin{array}{rcrcl}{#1} &: &{#2} &\to &{#3}\\&& {#4} &\mapsto &{#5}\end{array}}
\newcommand{\hypgeo}{\vphantom{F}_2F_1}

\newcommand{\ytens}[3]{\overset{#2}{\underset{#3}{\mathbf{Y}}}\vcenter{\hbox{\scriptsize$#1$}}}
\DeclareMathOperator{\tr}{tr}

\begin{document}

\markboth{Manuel Hohmann}
{Spherical harmonic d-tensors}

%
\catchline{}{}{}{}{}
%

\title{Spherical harmonic d-tensors}

\author{Manuel Hohmann}

\address{Laboratory of Theoretical Physics, Institute of Physics, University of Tartu, W. Ostwaldi 1, 50411 Tartu, Estonia\\
\email{manuel.hohmann@ut.ee}}

\maketitle

\begin{history}
\received{(Day Month Year)}
\revised{(Day Month Year)}
\end{history}

\begin{abstract}
Tensor harmonics are a useful mathematical tool for finding solutions to differential equations which transform under a particular representation of the rotation group $\mathrm{SO}(3)$. The aim of this work is to make use of this tool also in the setting of Finsler geometry, or more general geometries on the tangent bundle, where the objects of relevance are d-tensors on the tangent bundle, or tensors in a pullback bundle, instead of ordinary tensors. For this purpose, we construct a set of d-tensor harmonics for spherical symmetry and show how these can be used for calculations in Finsler geometry.
\end{abstract}


\section{Introduction}\label{sec:intro}
It is commonly understood that problems in differential geometry and its applications in physics simplify if they exhibit any symmetries, such as spherical or planar symmetries, which are most common to appear in physics via the action of a corresponding symmetry group on some underlying space or spacetime manifold. Symmetries under such a group action are commonly encoded in geometric objects defined on these manifolds which are invariant under the group action, or transform under particular representations of the symmetry group. A common example is the hydrogen atom, whose wave function is expressed by a complex function on the space manifold \(M = \mathbf{R}^3\), and where the spherical symmetry of the potential allows to express the angular part of the eigenfunctions of the Hamiltonian in terms of spherical harmonics, which are the eigenfunctions of the angular momentum operator. Another example is given by electromagnetic and gravitational radiation from a point-like source, which is described by tensor fields, and whose multipole expansion can be described by suitable representations of the rotation group on the space of tensor fields.

While the geometric objects in the aforementioned examples have in common that they are defined on the underlying manifold itself, one finds a different situation in Finsler and spray geometry. In these cases one rather deals with geometric objects defined on the tangent bundle of the underlying manifold. In the most simple case, such as that of a Finsler function, the relevant object is simply a real or complex function on the tangent bundle. Another very commonly encountered class of objects is given by d-tensors, which were introduced in~\cite{Miron:1994}. These objects have in common that they have a well-defined transformation behavior under the action of a symmetry group on the underlying base manifold via diffeomorphisms. This naturally raises the question for suitable bases on the spaces containing these objects, which can be decomposed into irreducible representations of the symmetry group, in analogy to the well-known spherical harmonics.

The aim of this article is to construct a suitable generalization of spherical harmonics to the space of d-tensors of arbitrary rank over the manifold \(M = \mathbf{R}^3\), which allows for a decomposition into irreducible representations of the rotation group. Note that this in particular includes the case of d-tensors of rank \(0\), which are simply functions on the tangent bundle, and which will be the starting point of our construction. We essentially follow the same steps as done for tensor spherical harmonics~\cite{James:1976}, and so the d-tensors we construct will inherit several properties of the tensor spherical harmonics on which our construction is based.

The article is structured as follows. We start with a brief review of the relevant mathematical notions and introduce the necessary objects for the description of spherical symmetry in section~\ref{sec:pre}. We then construct the spherical harmonic d-tensors in two steps. The most simple case of rank \(0\), which is simply given by functions on the tangent bundle \(TM\), will be discussed in section~\ref{sec:scalar}. This will turn out to be a necessary building block for the case of general rank discussed in section~\ref{sec:dtensor}. As an illustrative example, we calculate the Finsler metric for the most general spherically symmetric Finsler space in terms of spherical harmonic d-tensors in section~\ref{sec:app}. We end with a conclusion in section~\ref{sec:conclusion}.

\section{Preliminaries}\label{sec:pre}
Before we show our construction of spherical harmonic d-tensors, we clarify a few necessary notions. We start with the definition of d-tensors using the pullback formalism in section~\ref{ssec:pullback}. We then show how these d-tensors transform under diffeomorphisms of the base manifold in section~\ref{ssec:sym}. In order to be able to discuss spherical symmetry, we introduce suitable coordinates on the tangent bundle in section~\ref{ssec:coord}. Finally, in section~\ref{ssec:spher} we display the generators of rotations in our chosen coordinates.

\subsection{Pullback bundle formalism and d-tensors}\label{ssec:pullback}
One of the most important classes of objects in Finsler or spray geometry is given by \emph{distinguished tensors}, or simply d-tensors. These can be defined in different ways, the most common definition being as tensors over the tangent bundle \(TM\) of a manifold \(M\), whose components are purely horizontal with respect to some given splitting \(TTM = HTM \oplus VTM\) of the double tangent bundle and a corresponding splitting of \(T^*TM\). For our purposes, however, a different definition in terms of pullback bundles~\cite{Szilasi:2003} is more convenient. For this purpose we introduce the pullback bundle \(\pi: PM \to TM\), where \(PM = TM \times_M TM\) is a fibered product and \(\pi\) is the projection onto the first factor of this product. It is a vector bundle over \(TM\), whose fibers are isomorphic to the fibers of \(TM\), and whose dual is given by \(P^*M = TM \times_M T^*M\).

The pullback bundle is closely related to the tangent bundle \(\tau: TM \to M\) and the double tangent bundle \(\varpi: TTM \to TM\). This can be seen after defining the two maps
\begin{equation}
\func{\mathbf{i}}{PM}{TTM}{(v,w)}{\left.\frac{d}{dt}(v + tw)\right|_{t = 0}}
\end{equation}
and
\begin{equation}
\func{\mathbf{j}}{TTM}{PM}{\xi}{\left(\varpi(\xi), \tau_*(\xi)\right)}\,,
\end{equation}
called the \emph{vertical} and \emph{horizontal morphism}, respectively. One can show that they form an exact sequence
\begin{equation}\label{eq:exseq}
0 \rightarrow PM \xrightarrow{\mathbf{i}} TTM \xrightarrow{\mathbf{j}} PM \rightarrow 0
\end{equation}
of vector bundle morphisms over \(TM\).

From the pullback bundle and its dual one can now construct the tensor bundles
\begin{equation}
P^r_sM = \underbrace{PM \otimes \cdots \otimes PM}_{r \text{ times}} \otimes \underbrace{P^*M \otimes \cdots \otimes P^*M}_{s \text{ times}}\,,
\end{equation}
which are again bundles over \(TM\). We then call a \emph{d-tensor} of rank \((r,s)\) a section of the corresponding bundle \(\pi^r_s: P^r_sM \to TM\).

Given coordinates \((x^a)\) on \(M\), one has a coordinate basis \((\partial_a)\) on \(TM\), which allows to write any tangent vector \(v \in T_xM\) in the form \(v = \dot{x}^a\partial_a\), where we introduced coordinates \((\dot{x}^a)\) on \(T_xM\). This yields a set of induced coordinates \((x^a, \dot{x}^a)\) on \(TM\). Further, recall that the fiber \(P_vM\) of \(PM\) over \(v \in TM\) is canonically isomorphic to the fiber \(T_{\tau(v)}M\). One may therefore canonically lift the basis vector fields \(\partial_a\) on \(M\) to sections of \(PM\), which likewise have the property to form a basis at each \(v \in TM\). The same holds for \(P^*M\) and the basis covector fields \(dx^a\). This allows us to write any d-tensor \(A\) of rank \((r,s)\) in components as
\begin{equation}
A(x,\dot{x}) = A^{a_1 \cdots a_r}{}_{b_1 \cdots b_s}(x,\dot{x})\partial_{a_1} \otimes \cdots \otimes \partial_{a_r} \otimes dx^{b_1} \otimes \cdots \otimes dx^{b_s}\,.
\end{equation}
This component expression is also helpful for calculating the transformation behavior under symmetry transformations, as we discuss next.

\subsection{Transformation of d-tensors under diffeomorphisms}\label{ssec:sym}
We now briefly review the transformation behavior of d-tensors under finite and infinitesimal diffeomorphisms from the manifold \(M\) to itself. Let \(\varphi: M \to M\) be a diffeomorphism. Its differential \(\varphi_*: TM \to TM\) is a diffeomorphism from \(TM\) to itself. This lift of \(\varphi\) to the tangent bundle is a functorial lift; it shows that \(TM\) is an example of a natural bundle~\cite{Kolar:1993}. In induced coordinates \((x^a, \dot{x}^a)\) we may define \(x' = \varphi(x)\) and find that \(\varphi_*\) is expressed as
\begin{equation}
\dot{x}'^a = \frac{\partial x'^a}{\partial x^b}\dot{x}^b\,.
\end{equation}
Given a d-tensor \(A\) of rank \((r,s)\), we may calculate its pullback \(\varphi^*A\) as~\cite{Tashiro:1959}
\begin{equation}
(\varphi^*A)^{a_1 \cdots a_r}{}_{b_1 \cdots b_s}(x,\dot{x}) = A^{c_1 \cdots c_r}{}_{d_1 \cdots d_s}(x',\dot{x}')\frac{\partial x^{a_1}}{\partial x'^{c_1}} \cdots \frac{\partial x^{a_r}}{\partial x'^{c_r}}\frac{\partial x'^{d_1}}{\partial x^{b_1}} \cdots \frac{\partial x'^{d_s}}{\partial x^{b_s}}\,.
\end{equation}
If we are given a one-parameter group \(t \mapsto \varphi_t\) of diffeomorphisms instead, which is generated by the vector field
\begin{equation}
\func{\xi}{M}{TM}{x}{\left.\frac{d}{dt}\varphi_t(x)\right|_{t = 0}}\,,
\end{equation}
then we obtain the Lie derivative of a d-tensor as
\begin{equation}
\begin{split}
(\mathcal{L}_{\xi}A)^{a_1 \cdots a_r}{}_{b_1 \cdots b_s} &= \left.\frac{d}{dt}(\varphi^*A)^{a_1 \cdots a_r}{}_{b_1 \cdots b_s}\right|_{t = 0}\\
&= \xi^c\partial_cA^{a_1 \cdots a_r}{}_{b_1 \cdots b_s} + \dot{x}^d\partial_d\xi^c\dot{\partial}_cA^{a_1 \cdots a_r}{}_{b_1 \cdots b_s}\\
&\phantom{=}- \xi^{a_1}\partial_cA^{ca_2 \cdots a_r}{}_{b_1 \cdots b_s} - \cdots - \xi^{a_r}\partial_cA^{a_1 \cdots a_{r-1}c}{}_{b_1 \cdots b_s}\\
&\phantom{=}+ \xi^c\partial_{b_1}A^{a_1 \cdots a_r}{}_{cb_2 \cdots b_s} + \cdots + \xi^c\partial_{b_s}A^{a_1 \cdots a_r}{}_{b_1 \cdots b_{s-1}c}\,.
\end{split}
\end{equation}
The vector field \(\xi^c\partial_c + \dot{x}^d\partial_d\xi^c\dot{\partial}_c\) on \(TM\) generating the one-parameter diffeomorphism group \(\varphi_{t*}\) is called the complete lift~\cite{Yano:1973} of \(\xi\).

\subsection{Coordinates}\label{ssec:coord}
In this article we make use of different sets of coordinates on \(TM\). We start by introducing Cartesian coordinates \((x^a, a = 1, \ldots, 3)\) on \(M\). These induce a set of canonical coordinate basis vector fields, which we denote by \(\partial_a\), and which allow us to introduce coordinates \((\dot{x}^a, a = 1, \ldots, 3)\) as \(\dot{x}^a\partial_a\) on every fiber of \(TM\), and by restriction also on \(TM\). The corresponding coordinate basis of \(TTM\) will be denoted by \((\partial_a, \dot{\partial}_a)\).

Starting from the Cartesian coordinates, we define two more sets of coordinates on \(TM\). The first set \((r, \theta, \phi, \bar{r}, \alpha, \beta)\), which we call co-rotated spherical coordinates, is defined by
\begin{subequations}
\begin{align}
\left(\begin{array}{c}
x^1\\
x^2\\
x^3
\end{array}\right) &= \left(\begin{array}{ccc}
\cos\phi & -\sin\phi & 0\\
\sin\phi & \cos\phi & 0\\
0 & 0 & 1
\end{array}\right) \cdot \left(\begin{array}{ccc}
\cos\theta & 0 & \sin\theta\\
0 & 1 & 0\\
-\sin\theta & 0 & \cos\theta
\end{array}\right) \cdot \left(\begin{array}{c}
0\\
0\\
r
\end{array}\right)\\
\left(\begin{array}{c}
\dot{x}^1\\
\dot{x}^2\\
\dot{x}^3
\end{array}\right) &= \left(\begin{array}{ccc}
\cos\phi & -\sin\phi & 0\\
\sin\phi & \cos\phi & 0\\
0 & 0 & 1
\end{array}\right) \cdot \left(\begin{array}{ccc}
\cos\theta & 0 & \sin\theta\\
0 & 1 & 0\\
-\sin\theta & 0 & \cos\theta
\end{array}\right)\nonumber\\
&\phantom{=}\cdot \left(\begin{array}{ccc}
\cos\beta & -\sin\beta & 0\\
\sin\beta & \cos\beta & 0\\
0 & 0 & 1
\end{array}\right) \cdot \left(\begin{array}{ccc}
\cos\alpha & 0 & \sin\alpha\\
0 & 1 & 0\\
-\sin\alpha & 0 & \cos\alpha
\end{array}\right) \cdot \left(\begin{array}{c}
0\\
0\\
\bar{r}
\end{array}\right)
\end{align}
\end{subequations}
The second set \((r, \theta, \phi, \bar{\rho}, \bar{z}, \beta)\) will be called co-rotated cylindrical coordinates and defined as
\begin{subequations}
\begin{align}
\left(\begin{array}{c}
x^1\\
x^2\\
x^3
\end{array}\right) &= \left(\begin{array}{ccc}
\cos\phi & -\sin\phi & 0\\
\sin\phi & \cos\phi & 0\\
0 & 0 & 1
\end{array}\right) \cdot \left(\begin{array}{ccc}
\cos\theta & 0 & \sin\theta\\
0 & 1 & 0\\
-\sin\theta & 0 & \cos\theta
\end{array}\right) \cdot \left(\begin{array}{c}
0\\
0\\
r
\end{array}\right)\\
\left(\begin{array}{c}
\dot{x}^1\\
\dot{x}^2\\
\dot{x}^3
\end{array}\right) &= \left(\begin{array}{ccc}
\cos\phi & -\sin\phi & 0\\
\sin\phi & \cos\phi & 0\\
0 & 0 & 1
\end{array}\right) \cdot \left(\begin{array}{ccc}
\cos\theta & 0 & \sin\theta\\
0 & 1 & 0\\
-\sin\theta & 0 & \cos\theta
\end{array}\right) \cdot \left(\begin{array}{c}
\bar{\rho}\cos\beta\\
\bar{\rho}\sin\beta\\
\bar{z}
\end{array}\right)
\end{align}
\end{subequations}
They are obviously related to each other by
\begin{equation}\label{eq:sphercyltrans}
\bar{\rho} = \bar{r}\sin\alpha\,, \quad
\bar{z} = \bar{r}\cos\alpha\,,
\end{equation}
similar to the usual spherical and cylindrical coordinates. The usefulness of these coordinates will become apparent later.

We finally remark that by introducing Cartesian coordinates on \(M\) together with bases \((\partial_a)\) of \(TM\) and \((dx^a)\) of \(T^*M\) we have also introduced bases of the corresponding pullback bundles over \(TM\), which are simply obtained by pullback via the fiber isomorphisms. We will use the same notation for these bases. Hence, we have also defined bases for d-tensors, and we will use them throughout the remainder of this article.

\subsection{Spherical symmetry}\label{ssec:spher}
We now use the previously defined coordinates for the description of spherical symmetry, i.e., symmetry under rotations. Recall that rotations in Euclidean geometry can be defined as diffeomorphism that keep the origin fixed and preserve distances measured with the canonical Euclidean metric \(\delta_{ab} = \mathrm{diag}(1,1,1)\). For our purpose it is sufficient to consider the generating vector fields of rotations, which in Cartesian coordinates take the simple forms
\begin{equation}
\mathbf{r}_1 = -x^2\partial_3 + x^3\partial_2\,, \quad \mathbf{r}_2 = -x^3\partial_1 + x^1\partial_3\,, \quad \mathbf{r}_3 = -x^1\partial_2 + x^2\partial_1\,.
\end{equation}
In order to apply them to d-tensors and functions on \(TM\), we also need their canonical lifts. In Cartesian induced coordinates they are given by
\begin{subequations}
\begin{align}
\hat{\mathbf{r}}_1 &= -x^2\partial_3 + x^3\partial_2 - \dot{x}^2\dot{\partial}_3 + \dot{x}^3\dot{\partial}_2\,,\\
\hat{\mathbf{r}}_2 &= -x^3\partial_1 + x^1\partial_3 - \dot{x}^3\dot{\partial}_1 + \dot{x}^1\dot{\partial}_3\,,\\
\hat{\mathbf{r}}_3 &= -x^1\partial_2 + x^2\partial_1 - \dot{x}^1\dot{\partial}_2 + \dot{x}^2\dot{\partial}_1\,.
\end{align}
\end{subequations}
These expressions are still simple, but not the most useful for the construction of harmonics. We therefore express the canonical lifts in the other sets of coordinates we have introduced, and find that they take the same form
\begin{subequations}
\begin{align}
\hat{\mathbf{r}}_1 &= \sin\phi\partial_{\theta} + \frac{\cos\phi}{\tan\theta}\partial_{\phi} - \frac{\cos\phi}{\sin\theta}\partial_{\beta}\,,\\
\hat{\mathbf{r}}_2 &= -\cos\phi\partial_{\theta} + \frac{\sin\phi}{\tan\theta}\partial_{\phi} - \frac{\sin\phi}{\sin\theta}\partial_{\beta}\,,\\
\hat{\mathbf{r}}_3 &= -\partial_{\phi}
\end{align}
\end{subequations}
in both co-rotated spherical and co-rotated cylindrical coordinates. Note that they act only on the coordinates \(\theta, \phi, \beta\) and leave the other coordinates invariant. This will significantly simplify our task of constructing harmonics in the following sections. We finally define the operators \(\mathcal{R}_j\), which act on d-tensors \(A\) as
\begin{equation}
\mathcal{R}_jA = i\mathcal{L}_{\mathbf{r}_j}A\,,
\end{equation}
and we will use this notation throughout the remainder of this article.

\section{Tangent bundle spherical harmonics}\label{sec:scalar}
We now continue with the construction of functions on the tangent bundle \(TM\), which form irreducible representations of the algebra of rotation operators \(\mathcal{R}_j\) introduced in the previous section. This will be done in several steps. We start by enlarging the algebra by introducing another set of operators in section~\ref{ssec:corot}. We then choose a particular set of commuting operators, and construct their eigenfunctions in section~\ref{ssec:eigen}. This will lead us to the definition of tangent bundle spherical harmonics in section~\ref{ssec:scalharm}. We list a few special cases in section~\ref{ssec:scalspecial}, and discuss their properties in section~\ref{ssec:scalprop}.

\subsection{Co-rotation operators}\label{ssec:corot}
We start our discussion of harmonic functions by introducing another set of operators, which act on functions on the (slit) tangent bundle \(TM\). For this purpose we introduce the vector fields
\begin{subequations}
\begin{align}
\mathbf{b}_1 &= \sin\beta\partial_{\theta} + \frac{\cos\beta}{\tan\theta}\partial_{\beta} - \frac{\cos\beta}{\sin\theta}\partial_{\phi}\,,\\
\mathbf{b}_2 &= -\cos\beta\partial_{\theta} + \frac{\sin\beta}{\tan\theta}\partial_{\beta} - \frac{\sin\beta}{\sin\theta}\partial_{\phi}\,,\\
\mathbf{b}_3 &= -\partial_{\beta}
\end{align}
\end{subequations}
on \(TM\). Note that their expressions are identical in co-rotated spherical and co-rotated cylindrical coordinates, so that we do not have to distinguish between these two sets of coordinates at this point. It is important to note that they are \emph{not} the complete lifts of vector fields on the base manifold \(M\), and so they cannot be applied to d-tensors as discussed in section~\ref{ssec:sym}, but only on functions (or ordinary tensor fields) on \(TM\). Their action on functions \(f \in C^{\infty}\left(TM\right)\) is given by the usual Lie derivative, and allows us to define the operators
\begin{equation}
\mathcal{B}_jf = i\mathcal{L}_{\mathbf{b}_j}f\,,
\end{equation}
which we call \emph{co-rotation operators}. These operators satisfy the algebra relations
\begin{equation}
[\mathcal{B}_j,\mathcal{B}_k] = i\epsilon_{jkl}\mathcal{B}_l\,, \quad [\mathcal{B}_j,\mathcal{R}_k] = 0\,,
\end{equation}
where in the latter the restriction of \(\mathcal{R}_j\) to functions on \(TM\) is understood. We further define the operators
\begin{equation}
\mathcal{B}_{\pm} = \mathcal{B}_1 \pm i\mathcal{B}_2\,, \quad \mathcal{B}_z = \mathcal{B}_3\,, \quad \mathcal{B}^2 = \mathcal{B}_1^2 + \mathcal{B}_2^2 + \mathcal{B}_3^2 = \mathcal{R}^2\,.
\end{equation}
They satisfy the algebra relations
\begin{equation}
[\mathcal{B}_z,\mathcal{B}_{\pm}] = \pm\mathcal{B}_{\pm}\,, \quad [\mathcal{B}_+,\mathcal{B}_-] = 2\mathcal{B}_z\,, \quad [\mathcal{B}_{\pm},\mathcal{B}^2] = [\mathcal{B}_z,\mathcal{B}^2] = 0\,.
\end{equation}
We see that the operators \(\mathcal{B}_j\) and \(\mathcal{R}_j\) together satisfy the algebra of rotations of a rigid body.

\subsection{Eigenvalues and eigenfunctions}\label{ssec:eigen}
We now construct irreducible representations of the algebra of rigid rotations spanned by the operators \(\mathcal{B}_j\) and \(\mathcal{R}_j\) above. Since \(\mathcal{R}^2, \mathcal{R}_z, \mathcal{B}_z\) mutually commute, we can find simultaneous eigenfunctions \(f\) which satisfy
\begin{equation}\label{eq:eigenfunc}
\mathcal{R}^2f = l(l + 1)f\,, \quad \mathcal{R}_zf = mf\,, \quad \mathcal{B}_zf = nf\,.
\end{equation}
For this purpose we write a function \(f \in C^{\infty}\left(TM\right)\) using a separation ansatz in the form
\begin{equation}
f(r, \bar{r}, \alpha, \beta, \theta, \phi) = \tilde{f}(r, \bar{r}, \alpha)\Theta(\theta)\Phi(\phi)B(\beta)
\end{equation}
in co-rotated spherical coordinates into a ``radial'' part \(\tilde{f}\) and three functions constituting the ``angular'' part. Note that operators we consider act only on the angular part, so that we could equally well use co-rotated cylindrical coordinates instead and introduce the separation ansatz
\begin{equation}
f(r, \bar{\rho}, \bar{z}, \beta, \theta, \phi) = \check{f}(r, \bar{\rho}, \bar{z})\Theta(\theta)\Phi(\phi)B(\beta)\,,
\end{equation}
where the functions \(\tilde{f}\) and \(\check{f}\) are related through the coordinate transformation~\eqref{eq:sphercyltrans} by
\begin{equation}
\tilde{f}(r, \bar{r}, \alpha) = \check{f}(r, \bar{r}\sin\alpha, \bar{r}\cos\alpha)\,.
\end{equation}
From this separation ansatz we then obtain the solutions
\begin{equation}
\Phi(\phi) = e^{im\phi}\,, \quad B(\beta) = e^{in\beta}
\end{equation}
for two of the eigenvalue equations. Since the coordinates \(\phi\) and \(\beta\) are $2\pi$-periodic, \(m\) and \(n\) must be integers. We can insert this solution into the equation for \(\mathcal{R}^2\) and obtain
\begin{equation}
\Theta''(\theta) + \frac{\Theta'(\theta)}{\tan\theta} + \left(\frac{2mn\cos\theta - m^2 - n^2}{\sin^2\theta} + l(l + 1)\right)\Theta(\theta) = 0\,.
\end{equation}
This equation can be solved more easily by introducing a new variable \(z = \cos\theta\) and a function \(Z(\cos\theta) = \Theta(\theta)\). The differential equation then reads
\begin{equation}
(1 - z^2)Z''(z) - 2zZ'(z) + \left(\frac{2mnz - m^2 - n^2}{1 - z^2} + l(l + 1)\right)Z(z) = 0\,.
\end{equation}
This equation has the general solution
\begin{multline}
Z(z) = \left(\frac{1 + z}{1 - z}\right)^{\frac{m + n}{2}}\bigg[C_1(1 - z)^m\hypgeo\left(m - l, l + m + 1; 1 + m - n; \frac{1 - z}{2}\right)\\
+ C_2(1 - z)^n\hypgeo\left(n - l, l + n + 1; 1 + n - m; \frac{1 - z}{2}\right)\bigg]
\end{multline}
with integration constants \(C_1, C_2\). Here we are interested in regular solutions on the interval \(z \in [-1,1]\). The regular solution is given by
\begin{multline}
Z(z) = C\sqrt{1 + z}^{m + n}\sqrt{1 - z}^{|m - n|}\\
\hypgeo\left(\max(m,n) - l, \max(m,n) + l + 1; |m - n| + 1; \frac{1 - z}{2}\right)
\end{multline}
with \(l \in \mathbb{N}\) and \(m, n \in \{-l, -l + 1, \ldots, l\}\), and \(C\) is an integration constant. After substituting back to \(\theta\) we thus obtain the solution
\begin{multline}
\Theta(\theta) = C'\cos^{m + n}\frac{\theta}{2}\sin^{|m - n|}\frac{\theta}{2}\\
\hypgeo\left(\max(m,n) - l, \max(m,n) + l + 1; |m - n| + 1; \sin^2\frac{\theta}{2}\right)\,,
\end{multline}
where \(l, m, n\) take the same values as above. Note that up to the integration constant the functions \(\Theta, \Phi, B\) are uniquely defined by the eigenvalue equations~\eqref{eq:eigenfunc} and the requirement that they are regular.

\subsection{Definition and explicit formula}\label{ssec:scalharm}
We can now define a family of functions
\begin{equation}
\begin{split}
\mathcal{Y}_{l,m,n}(\theta,\phi,\beta) &= N_{l,m,n}e^{im\phi}e^{in\beta}\cos^{m + n}\frac{\theta}{2}\sin^{|m - n|}\frac{\theta}{2}\\
&\phantom{=}\cdot \hypgeo\left(\max(m,n) - l, \max(m,n) + l + 1; |m - n| + 1; \sin^2\frac{\theta}{2}\right)\,,
\end{split}
\end{equation}
where the normalization constants
\begin{equation}
N_{l,m,n} = (-1)^{\max(m,n)}\frac{\sqrt{(2l + 1)}}{|m - n|!}\sqrt{\frac{(l - \min(m,n))!(l + \max(m,n))!}{(l - \max(m,n))!(l + \min(m,n))!}}
\end{equation}
are chosen so that
\begin{equation}
\int_{0}^{2\pi}\int_{0}^{2\pi}\int_{0}^{\pi}|\mathcal{Y}_{l,m,n}(\theta,\phi,\beta)|^2\sin\theta\,d\theta\,d\phi\,d\beta = 8\pi^2\,.
\end{equation}
We call these functions the tangent bundle spherical harmonics. Note that up to the choice of the normalization constants and notation they are the same as the Wigner d-matrix components, which are defined as
\begin{multline}
D^l_{m,n}(\phi, \theta, \beta) = (-1)^{m-n}\sqrt{(l+m)!(l-m)!(l+n)!(l-n)!}e^{-im\phi}e^{-in\beta}\\
\sum_s\frac{(-1)^s}{s!(l+n-s)!(l-m-s)!(m-n+s)!}\cos^{2l+n-m-2s}\frac{\theta}{2}\sin^{m-n+2s}\frac{\theta}{2}\,.
\end{multline}
One finds that \(\mathcal{Y}_{l,m,n} = (-1)^m\sqrt{2l + 1}D^l_{-m,-n}\).

\subsection{Special cases}\label{ssec:scalspecial}
We note that there are a number of special cases. For \(n = 0\) the functions do not depend on the tangent bundle coordinates, and reduce to the usual spherical harmonics
\begin{equation}
\mathcal{Y}_{l,m,0}(\theta,\phi,\beta) = \sqrt{4\pi}Y_{lm}(\theta,\phi)\,.
\end{equation}
It further follows from the symmetry \(\phi \leftrightarrow \beta\), \(m \leftrightarrow n\) that analogously holds
\begin{equation}
\mathcal{Y}_{l,0,n}(\theta,\phi,\beta) = \sqrt{4\pi}Y_{ln}(\theta,\beta)
\end{equation}
for \(m = 0\).

\subsection{Properties}\label{ssec:scalprop}
\subsubsection{Operator relations}
The harmonics satisfy the relations
\begin{equation}
\mathcal{R}^2\mathcal{Y}_{l,m,n} = l(l + 1)\mathcal{Y}_{l,m,n}\,, \quad \mathcal{R}_z\mathcal{Y}_{l,m,n} = m\mathcal{Y}_{l,m,n}\,, \quad \mathcal{B}_z\mathcal{Y}_{l,m,n} = n\mathcal{Y}_{l,m,n}\,.
\end{equation}
Functions with identical \(l\) are related by application of the ladder operators
\begin{subequations}
\begin{align}
\mathcal{R}_{\pm}\mathcal{Y}_{l,m,n} &= \sqrt{(l \mp m)(l \pm m + 1)}\mathcal{Y}_{l,m \pm 1,n}\,,\\
\mathcal{B}_{\pm}\mathcal{Y}_{l,m,n} &= \sqrt{(l \mp n)(l \pm n + 1)}\mathcal{Y}_{l,m,n \pm 1}\,.
\end{align}
\end{subequations}

\subsubsection{Complex conjugate}
The complex conjugate is given by\begin{equation}
\overline{\mathcal{Y}_{l,m,n}}(\theta,\phi,\beta) = (-1)^{m + n}\mathcal{Y}_{l,-m,-n}(\theta,\phi,\beta)\,.
\end{equation}

\subsubsection{Orthogonality}
The harmonics are orthogonal,
\begin{equation}
\int_{0}^{2\pi}\int_{0}^{2\pi}\int_{0}^{\pi}\mathcal{Y}_{l,m,n}(\theta,\phi,\beta)\overline{\mathcal{Y}_{l',m',n'}}(\theta,\phi,\beta)\sin\theta\,d\theta\,d\phi\,d\beta = 8\pi^2\delta_{ll'}\delta_{mm'}\delta_{nn'}\,.
\end{equation}

\subsubsection{Product rule}
The product of two harmonics can be written as
\begin{equation}
\begin{split}
\mathcal{Y}_{l,m,n}\mathcal{Y}_{l',m',n'} &= \sum_{l''}\sqrt{\frac{(2l + 1)(2l' + 1)}{2l'' + 1}}C^{l,l',l''}_{m,m',m+m'}C^{l,l',l''}_{n,n',n+n'}\mathcal{Y}_{l'',m+m',n+n'}\\
&= (-1)^{m+m'+n+n'}\sum_{l''}\sqrt{(2l + 1)(2l' + 1)(2l'' + 1)}\\
&\phantom{=}\cdot \left(\begin{array}{ccc}
l & l' & l''\\
m & m' & -m - m'
\end{array}\right)\left(\begin{array}{ccc}
l & l' & l''\\
n & n' & -n - n'
\end{array}\right)\mathcal{Y}_{l'',m+m',n+n'}\,,
\end{split}
\end{equation}
where \(C^{l,l',l''}_{m,m',m''}\) denotes the Clebsch-Gordan coefficients and the terms in the sum are non-vanishing only for
\begin{equation}
\max(|l - l'|, |m + m'|, |n + n'|) \leq l'' \leq l + l'\,.
\end{equation}

\section{Spherical harmonic d-tensors}\label{sec:dtensor}
Using the tangent bundle spherical harmonics introduced in the previous section, we now finally come to the construction of spherically harmonic d-tensors. Again we proceed in several steps. We start by choosing suitable basis vectors and covector, which form a representation of the rotation algebra in section~\ref{ssec:basis}. From these and their tensor products we then construct the spherical harmonic d-tensors in section~\ref{ssec:tensharm}. Some special cases are listed in section~\ref{ssec:tensspecial}, and their properties are discussed in section~\ref{ssec:tensprop}.

\subsection{Basis vectors and covectors}\label{ssec:basis}
We start our construction by examining the action of the rotation operators \(\mathcal{R}_a\) on the coordinate basis elements \(\partial_a\) and \(dx^a\) of the pullback bundles \(TM \times_M TM\) and \(TM \times_M T^*M\) derived from the Cartesian coordinates in section~\ref{ssec:coord}. Starting with the former, one finds that they are given by
\begin{equation}
\mathcal{R}_a\partial_b = i\epsilon_{abc}\partial_c\,,
\end{equation}
as usual for an oriented right-handed vector basis. In particular, it follows that they satisfy the relations
\begin{equation}
\mathcal{R}^2\partial_a = 2\partial_a\,, \quad
\mathcal{R}_z\partial_1 = i\partial_2\,, \quad
\mathcal{R}_z\partial_2 = -i\partial_1\,, \quad
\mathcal{R}_z\partial_3 = 0\,.
\end{equation}
This motivates the definition of the basis d-tensors
\begin{equation}
\mathbf{e}_0 = \bar{\partial}_3\,, \quad \mathbf{e}_1 = -\frac{\bar{\partial}_1 + i\bar{\partial}_2}{\sqrt{2}}\,, \quad \mathbf{e}_{-1} = \frac{\bar{\partial}_1 - i\bar{\partial}_2}{\sqrt{2}}\,.
\end{equation}
The d-tensors defined above satisfy the relations
\begin{equation}\label{eq:basvectrel}
\mathcal{R}^2\mathbf{e}_m = 2\mathbf{e}_m\,, \quad \mathcal{R}_z\mathbf{e}_m = m\mathbf{e}_m\,, \quad \mathcal{R}_{\pm}\mathbf{e}_m = \sqrt{(1 \mp m)(2 \pm m)}\mathbf{e}_{m \pm 1}\,.
\end{equation}
Note that the same construction can be applied to the covector basis \((dx^a)\). In this case we define
\begin{equation}
\mathbf{e}^0 = dx^3\,, \quad \mathbf{e}^1 = -\frac{dx^1 + idx^2}{\sqrt{2}}\,, \quad \mathbf{e}^{-1} = \frac{dx^1 - idx^2}{\sqrt{2}}
\end{equation}
They satisfy the same relations~\eqref{eq:basvectrel} as the corresponding vector basis elements.

\subsection{Definition and explicit formula}\label{ssec:tensharm}
We now recursively construct the rank \(k\) d-tensors from the rank \(1\) basis tensors as follows. We start with the rank \(0\) tensors, which are simply given by
\begin{equation}
\ytens{l}{m}{n} = \mathcal{Y}_{l,m,n}\,.
\end{equation}
Note that we have vertically centered the index \(l\) for this scalar, the reason of which will become clear in the following. From this we construct the vectors
\begin{equation}
\ytens{l'}{m}{n}{}_{l} = (-1)^{l - m}\sqrt{2l + 1}\sum_{m',\mu}\left(\begin{array}{ccc}
l & l' & 1\\
m & -m' & -\mu
\end{array}\right)\mathcal{Y}_{l',m',n}\mathbf{e}_{\mu}\,,
\end{equation}
and analogously the covectors
\begin{equation}
\ytens{l'}{m}{n}{}^{l} = (-1)^{l - m}\sqrt{2l + 1}\sum_{m',\mu}\left(\begin{array}{ccc}
l & l' & 1\\
m & -m' & -\mu
\end{array}\right)\mathcal{Y}_{l',m',n}\mathbf{e}^{\mu}\,,
\end{equation}
where the position of the new index reflects the tensor type. Higher order tensors are then constructed using the recursion formula
\begin{equation}
\ytens{l_0}{m}{n}{}_{l_1 \cdots l_k} = (-1)^{l_k - m}\sqrt{2l_k + 1}\sum_{m',\mu}\left(\begin{array}{ccc}
l_k & l_{k-1} & 1\\
m & -m' & -\mu
\end{array}\right)\ytens{l_0}{m'}{n}{}_{l_1 \cdots l_{k-1}} \otimes \mathbf{e}_{\mu}
\end{equation}
and analogously for \(\mathbf{e}^m\) and mixed tensors. Here the appearance of the Wigner $3j$-symbols imply that the indices must obey the conditions
\begin{equation}
l_0 = 0, 1, \ldots\,, \quad l_i = |l_{i - 1} - 1|, \ldots, l_{i - 1} + 1\,, \quad m = -l_k, \ldots, l_k\,, \quad n = -l_0, \ldots, l_0\,,
\end{equation}
since otherwise they would vanish identically. Note that any tensor of rank \(k\) now carries \(k\) indices \(l_1, \ldots, l_k\) which are either in lower or upper position, and whose position reflects the position of the indices on the basis elements. For example, a tensor \(\ytens{l_0}{m}{n}_{l_1}{}^{l_2}\) would have a component expression of the form \(A^a{}_b\partial_a \otimes dx^b\). The reason for this choice is that the harmonic tensors form an adapted basis of the space of d-tensors, in analogy to the coordinate basis elements such as \(\partial_a \otimes dx^b\), and so should carry the indices in the same positions as it is the case for the coordinate basis.

From the recursive definition one easily derives the explicit formula
\begin{multline}
\ytens{l_0}{m_k}{n}{}_{l_1 \cdots l_k} = \sum_{\substack{m_0, \ldots, m_{k-1}\\\mu_1, \ldots, \mu_k}}\mathcal{Y}_{l_0,m_0,n}\mathbf{e}_{\mu_1} \otimes \ldots \otimes \mathbf{e}_{\mu_k}\\
\cdot \prod_{i = 1}^k(-1)^{l_i - m_i}\sqrt{2l_i + 1}\left(\begin{array}{ccc}
l_i & l_{i-1} & 1\\
m_i & -m_{i-1} & -\mu_i
\end{array}\right)
\end{multline}
and analogously for \(\mathbf{e}^m\) and mixed tensors, by raising the indices \(\mu_i\) corresponding to the indices \(l_i\) for \(i = 1, \ldots, k\).

\subsection{Special cases}\label{ssec:tensspecial}
A few special cases are worth mentioning. First note that the basis vectors and covectors are recovered as
\begin{equation}
\mathbf{e}_m = \ytens{0}{m}{0}_1\,, \quad
\mathbf{e}^m = \ytens{0}{m}{0}^1\,.
\end{equation}
The Kronecker tensor is given by
\begin{equation}
\boldsymbol{\delta} = \partial_a \otimes dx^a = \sum_m\overline{\mathbf{e}_m} \otimes \mathbf{e}^m = -\sqrt{3}\ytens{0}{0}{0}{}_1{}^0\,.
\end{equation}
Finally, the totally antisymmetric tensor is given by
\begin{equation}
\boldsymbol{\epsilon} = \epsilon_{abc}dx^a \otimes dx^b \otimes dx^c = i\sqrt{6}\ytens{0}{0}{0}{}^{1\,1\,0}\,.
\end{equation}
Note that the latter two have \(l_k = 0\), and thus are invariant under rotation.

\subsection{Properties}\label{ssec:tensprop}
\subsubsection{Operator relations}
The harmonic d-tensors satisfy
\begin{gather}
\mathcal{R}^2\ytens{l_0}{m}{n}{}_{l_1 \cdots l_k} = l_k(l_k + 1)\ytens{l_0}{m}{n}{}_{l_1 \cdots l_k}\,, \quad \mathcal{R}_z\ytens{l_0}{m}{n}{}_{l_1 \cdots l_k} = m\ytens{l_0}{m}{n}{}_{l_1 \cdots l_k}\,,\nonumber\\
\mathcal{R}_{\pm}\ytens{l_0}{m}{n}{}_{l_1 \cdots l_k} = \sqrt{(l_k \mp m)(l_k \pm m + 1)}\ytens{l_0}{m \pm 1}{n}{}_{l_1 \cdots l_k}\,.
\end{gather}

\subsubsection{Complex conjugate}
The complex conjugate is given by
\begin{equation}
\overline{\ytens{l_0}{m}{n}{}_{l_1 \cdots l_k}} = (-1)^{l_0 + l_k + m + n + k}\ytens{l_0}{-m}{-n}{}_{l_1 \cdots l_k}\,.
\end{equation}

\subsubsection{Orthogonality}
The spherical harmonic d-tensors satisfy
\begin{equation}
\int_{0}^{2\pi}\int_{0}^{2\pi}\int_{0}^{\pi}\left\langle\ytens{l_0}{m}{n}{}_{l_1 \cdots l_k}, \ytens{l_0'}{m'}{n'}{}^{l_1 \cdots l_k}\right\rangle(\theta,\phi,\beta)\sin\theta\,d\theta\,d\phi\,d\beta = 8\pi^2\delta_{mm'}\delta_{nn'}\prod_{i = 0}^k\delta_{l_il_i'}\,,
\end{equation}
where
\begin{equation}
\langle A,B \rangle = \overline{A_{a_1 \cdots a_k}}B^{a_k \cdots a_1}\,.
\end{equation}

\subsubsection{Permutation of tensor indices}
The easiest relation one may derive is the transpose of tensors of rank \(2\). From the relation between $3j$-symbols and $6j$-symbols follows
\begin{equation}
\left(\ytens{l_0}{m}{n}{}_{l_1l_2}\right)^t = \sum_l(-1)^{l + l_1}\sqrt{2l + 1}\sqrt{2l_1 + 1}\left\{\begin{array}{ccc}
l_0 & l_1 & 1\\
l_2 & l & 1
\end{array}\right\}\ytens{l_0}{m}{n}{}_{ll_2}\,.
\end{equation}
From this formula one can then derive the transposition of neighboring indices. It can be constructed by making use of the recursive definition of the harmonic d-tensors. Denoting
\begin{multline}
\left(\mathbf{e}_{m_1} \otimes \ldots \otimes \mathbf{e}_{m_k}\right)^{t(i,j)} =\\
\mathbf{e}_{m_1} \otimes \ldots \otimes \mathbf{e}_{m_{i - 1}} \otimes \mathbf{e}_{m_j} \otimes \mathbf{e}_{m_{i + 1}} \otimes \ldots \otimes \mathbf{e}_{m_{j - 1}} \otimes \mathbf{e}_{m_i} \otimes \mathbf{e}_{m_{j + 1}} \otimes \ldots \otimes \mathbf{e}_{m_k}
\end{multline}
the transposition of indices \(i, j\) with \(i < j\), we find
\begin{equation}\label{eq:permneigh}
\left(\ytens{l_0}{m}{n}{}_{l_1 \cdots l_k}\right)^{t(i,i+1)} = \sum_l(-1)^{l + l_i}\sqrt{2l + 1}\sqrt{2l_i + 1}\left\{\begin{array}{ccc}
l_{i - 1} & l_i & 1\\
l_{i + 1} & l & 1
\end{array}\right\}\ytens{l_0}{m}{n}{}_{l_1 \cdots l_{i - 1}ll_{i + 1} \cdots l_k}\,.
\end{equation}
This can finally be generalized to arbitrary index permutations. Note that an arbitrary permutation of indices can be decomposed into a composition of transpositions of neighboring indices, and the aforementioned formula is applied recursively. The result is rather lengthy, and so we omit it here for brevity.

\subsubsection{Contraction}
Again we start with the most simple case given by tensors or rank \(2\). From the traces
\begin{equation}
\tr\left(\mathbf{e}_{m_1} \otimes \mathbf{e}^{m_2}\right) = \tr(\mathbf{e}^{m_1} \otimes \mathbf{e}_{m_2}) = (-1)^{m_1}\delta_{m_1,-m_2}
\end{equation}
and the properties of the $3j$-symbols follows:
\begin{equation}
\tr\ytens{l_0}{m}{n}{}_{l_1}{}^{l_2} = \tr\ytens{l_0}{m}{n}{}^{l_1}{}_{l_2} = (-1)^{l_0 - l_1}\sqrt{\frac{2l_1 + 1}{2l_0 + 1}}\delta_{l_0l_2}\mathcal{Y}_{l_0,m,n}\,.
\end{equation}
This formula generalizes directly to contractions over neighboring indices. Denoting this contraction by \(\tr_{i,j}\), we find
\begin{multline}
\tr_{i,i+1}\ytens{l_0 \cdots l_{i - 1}}{m}{n}{}_{l_i}{}^{l_{i + 1}}\vcenter{\hbox{\scriptsize$l_{i + 2} \cdots l_k$}} = \tr_{i,i+1}\ytens{l_0 \cdots l_{i - 1}}{m}{n}{}^{l_i}{}_{l_{i + 1}}\vcenter{\hbox{\scriptsize$l_{i + 2} \cdots l_k$}}\\
= (-1)^{l_{i + 1} - l_i}\sqrt{\frac{2l_i + 1}{2l_{i + 1} + 1}}\delta_{l_{i - 1}l_{i + 1}}\ytens{l_0 \cdots l_{i - 2}l_{i + 1} \cdots l_k}{m}{n}\,.
\end{multline}
Arbitrary contractions can then be calculated by first performing a permutation of indices by recursively applying the formula~\eqref{eq:permneigh} and then a contraction of neighboring indices.

\subsubsection{Tensor product}
The calculation of the tensor product is very lengthy, as it involves the evaluation of numerous $3j$-symbols. Nevertheless, it is possible to derive a general formula for the result, which is given by
\begin{multline}
\ytens{l_0}{m_k}{n}{}_{l_1 \cdots l_k} \otimes \ytens{l_0'}{m_{k'}'}{n'}{}_{l_1' \cdots l_{k'}'} = \sum_{\substack{l_0'',\ldots,l_{k+k'}''\\m_{k+k'}'',n''}}\left[\sum_{\substack{m_0,\ldots,m_{k-1}\\\mu_0,\ldots,\mu_{k-1}}}\sum_{\substack{m_0',\ldots,m_{k'-1}'\\\mu_0',\ldots,\mu_{k'-1}'}}\sum_{m_0'',\ldots,m_{k+k'-1}''}\right.\\
(-1)^{-m_0 - m_0' - n - n'}\left(\begin{array}{ccc}
l_0 & l_0' & l_0''\\
m_0 & m_0' & -m_0''
\end{array}\right)\left(\begin{array}{ccc}
l_0 & l_0' & l_0''\\
n & n' & -n''
\end{array}\right)\\
\cdot \prod_{i = 0}^{k-1}(-1)^{l_{i+1} + l_{i+1}'' - m_{i+1} - m_{i+1}''}\left(\begin{array}{ccc}
l_{i+1} & l_i & 1\\
m_{i+1} & -m_i & -\mu_i
\end{array}\right)\left(\begin{array}{ccc}
l_{i+k'+1}'' & l_{i+k'}'' & 1\\
m_{i+k'+1}'' & -m_{i+k'}'' & -\mu_i
\end{array}\right)\\
\cdot \prod_{i = 0}^{k' - 1}(-1)^{l_{i+1}' + l_{i+k+1}'' - m_{i+1}' - m_{i+k+1}''}\left(\begin{array}{ccc}
l_{i+1}' & l_i' & 1\\
m_{i+1}' & -m_i' & -\mu_i'
\end{array}\right)\left(\begin{array}{ccc}
l_{i+1}'' & l_i'' & 1\\
m_{i+1}'' & -m_i'' & -\mu_i'
\end{array}\right)\\
\cdot \left.\sqrt{\prod_{i = 0}^k(2l_i + 1)\prod_{i = 0}^{k'}(2l_i' + 1)\prod_{i = 0}^{k + k'}(2l_i'' + 1)}\right]\ytens{l_0''}{m_{k+k'}''}{n''}{}_{l_1'' \cdots l_{k+k'}''}\,,
\end{multline}
and where all sums have only a finite number of terms for which the $3j$-symbols are non-vanishing.

\subsubsection{Scalar product}
To calculate the scalar product of a vector and a covector, one may first calculate their tensor product, which follows from the general tensor product formula and takes the form
\begin{multline}
\ytens{l_0}{m}{n}{}_{l_1} \otimes \ytens{l_0'}{m'}{n'}{}^{l_1'} = \sum_{l_0'',l_1'',l_2''}(-1)^{m + m' + n + n'}\\
\cdot \sqrt{(2l_0 + 1)(2l_1 + 1)(2l_0' + 1)(2l_1' + 1)(2l_0'' + 1)(2l_1'' + 1)(2l_2'' + 1)}\\
\cdot \left(\begin{array}{ccc}
l_0'' & l_0' & l_0\\
-n - n' & n' & n
\end{array}\right)\left(\begin{array}{ccc}
l_2'' & l_1' & l_1\\
-m - m' & m' & m
\end{array}\right)\left\{\begin{array}{ccc}
1 & l_0 & l_1\\
l_0' & l_1'' & l_0''
\end{array}\right\}\left\{\begin{array}{ccc}
1 & l_0' & l_1'\\
l_1 & l_2'' & l_1''
\end{array}\right\}\ytens{l_0''}{m + m'}{n + n'}{}_{l_1''}{}^{l_2''}\,.
\end{multline}
By taking the trace we can read off the scalar product
\begin{multline}
\ytens{l_0}{m}{n}{}_{l_1} \cdot \ytens{l_0'}{m'}{n'}{}^{l_1'} = \sum_{l''}(-1)^{l_0 + l_1' + l'' + m + m' + n + n'}\\
\cdot \sqrt{(2l_0 + 1)(2l_1 + 1)(2l_0' + 1)(2l_1' + 1)(2l'' + 1)}\\
\cdot \left(\begin{array}{ccc}
l'' & l_0' & l_0\\
-n - n' & n' & n
\end{array}\right)\left(\begin{array}{ccc}
l'' & l_1' & l_1\\
-m - m' & m' & m
\end{array}\right)\left\{\begin{array}{ccc}
1 & l_0 & l_1\\
l'' & l_1' & l_0'
\end{array}\right\}\mathcal{Y}_{l'',m + m',n + n'}\,,
\end{multline}
where the two $6j$-symbols collapse into one by using their summation rules.

\subsubsection{Vertical differential}
The vertical differential \(\nabla^v\) of a harmonic d-tensor can most easily be written using co-rotated cylindrical coordinates \(r,\theta,\phi,\bar{\rho},\beta,\bar{z}\). For a function \(f(r,\bar{\rho},\bar{z})\) it takes the form
\begin{equation}
\begin{split}
\nabla^v[f\ytens{l_0}{m}{n}{}_{l_1 \cdots l_k}] &= dx^a \otimes \dot{\partial}_a[f\ytens{l_0}{m}{n}{}_{l_1 \cdots l_k}]\\
&= \left[\frac{1}{\sqrt{2}}\left(n\frac{f}{\bar{\rho}} - f_{\bar{\rho}}\right)\ytens{1}{0}{1}{}^0 + \frac{1}{\sqrt{2}}\left(n\frac{f}{\bar{\rho}} + f_{\bar{\rho}}\right)\ytens{1}{0}{-1}{}^0 - f_{\bar{z}}\ytens{1}{0}{0}{}^0\right] \otimes \ytens{l_0}{m}{n}{}_{l_1 \cdots l_k}\,.
\end{split}
\end{equation}
The vertical differential appears in various calculations in particular in Finsler geometry. A possible application will be shown in the following section.

\section{Application: spherically symmetric Finsler metric}\label{sec:app}
As an example, we now assume that our manifold \(M\) is equipped with a spherically symmetric Finsler Lagrangian \(L = L(r, \bar{\rho}, \bar{z})\) of homogeneity \(2\). We first calculate the momenta, given by the one-form
\begin{equation}
p_adx^a = \frac{1}{2}\bar{\partial}_aL\,dx^a = \frac{1}{2}\nabla^vL = -\frac{1}{2}L_{\bar{z}}\ytens{1}{0}{0}{}^0 - \frac{1}{2\sqrt{2}}L_{\bar{\rho}}\left(\ytens{1}{0}{1}{}^0 - \ytens{1}{0}{-1}{}^0\right)\,.
\end{equation}
Note that the appearing harmonic d-tensors have the property that \(l_k = 0\), which means that they are invariant under rotation, as one would expect, since we started from a Finsler Lagrangian with the same symmetry property. The same also holds for the Finsler metric
\begin{equation}
\begin{split}
g^L_{ab}dx^a \otimes dx^b &= \frac{1}{2}\bar{\partial}_a\bar{\partial}_bL\,dx^a \otimes dx^b = \frac{1}{2}\nabla^v\nabla^vL\\
&= -\frac{1}{2\sqrt{3}}\left(\frac{L_{\bar{\rho}}}{\bar{\rho}} + L_{\bar{\rho}\bar{\rho}} + L_{\bar{z}\bar{z}}\right)\ytens{0}{0}{0}{}^{1\,0} - \frac{1}{2\sqrt{6}}\left(\frac{L_{\bar{\rho}}}{\bar{\rho}} + L_{\bar{\rho}\bar{\rho}} - 2L_{\bar{z}\bar{z}}\right)\ytens{2}{0}{0}{}^{1\,0}\\
&\phantom{=}+ \frac{1}{2}L_{\bar{\rho}\bar{z}}\left(\ytens{2}{0}{1}{}^{1\,0} - \ytens{2}{0}{-1}{}^{1\,0}\right) + \frac{1}{4}\left(L_{\bar{\rho}\bar{\rho}} - \frac{L_{\bar{\rho}}}{\bar{\rho}}\right)\left(\ytens{2}{0}{2}{}^{1\,0} + \ytens{2}{0}{-2}{}^{1\,0}\right)\,.
\end{split}
\end{equation}
As a final illustration we also calculate its inverse. We remark that the derivation of a general formula for the inverse of a non-degenerate tensor of rank \(2\) is rather lengthy. However, one may use the tensor product and contraction formulas in order to derive a system of linear equations for the particular case we discuss here, and solve it explicitly. We then find that the inverse takes the form
\begin{equation}
\begin{split}
g^{L\,ab}\bar{\partial}_a \otimes \bar{\partial}_b &= \frac{2}{\sqrt{3}}\frac{L_{\bar{\rho}}\left(L_{\bar{\rho}\bar{\rho}} + L_{\bar{z}\bar{z}}\right) - \bar{\rho}\left(L_{\bar{\rho}\bar{z}}^2 - L_{\bar{\rho}\bar{\rho}}L_{\bar{z}\bar{z}}\right)}{L_{\bar{\rho}}\left(L_{\bar{\rho}\bar{z}}^2 - L_{\bar{\rho}\bar{\rho}}L_{\bar{z}\bar{z}}\right)}\ytens{0}{0}{0}{}_{1\,0}\\
&\phantom{=}- \frac{\sqrt{2}}{\sqrt{3}}\frac{L_{\bar{\rho}}\left(2L_{\bar{\rho}\bar{\rho}} - L_{\bar{z}\bar{z}}\right) + \bar{\rho}\left(L_{\bar{\rho}\bar{z}}^2 - L_{\bar{\rho}\bar{\rho}}L_{\bar{z}\bar{z}}\right)}{L_{\bar{\rho}}\left(L_{\bar{\rho}\bar{z}}^2 - L_{\bar{\rho}\bar{\rho}}L_{\bar{z}\bar{z}}\right)}\ytens{2}{0}{0}{}_{1\,0}\\
&\phantom{=}+ \frac{2L_{\bar{\rho}\bar{z}}}{L_{\bar{\rho}\bar{z}}^2 - L_{\bar{\rho}\bar{\rho}}L_{\bar{z}\bar{z}}}\left(\ytens{2}{0}{1}{}_{1\,0} - \ytens{2}{0}{-1}{}_{1\,0}\right)\\
&\phantom{=}- \frac{L_{\bar{\rho}}L_{\bar{z}\bar{z}} + \bar{\rho}\left(L_{\bar{\rho}\bar{z}}^2 - L_{\bar{\rho}\bar{\rho}}L_{\bar{z}\bar{z}}\right)}{L_{\bar{\rho}}\left(L_{\bar{\rho}\bar{z}}^2 - L_{\bar{\rho}\bar{\rho}}L_{\bar{z}\bar{z}}\right)}\left(\ytens{2}{0}{2}{}_{1\,0} + \ytens{2}{0}{-2}{}_{1\,0}\right)\,.
\end{split}
\end{equation}
One may then proceed with calculating further geometric objects, such as the Cartan tensor, in the same fashion. However, we stop at this point, as it should be clear now how these calculations are performed, and deriving all relevant quantities for a spherically symmetric Finsler space would exceed the scope of this article.

\section{Conclusion}\label{sec:conclusion}
We presented a set of d-tensors on the tangent bundle of a three-dimensional manifold, which decomposes into irreducible representations of the group of rotations acting on the underlying manifold. These d-tensors were constructed using suitable basis vectors and covectors as well as a set of spherical harmonic functions on the tangent bundle. We gave both a recursive definition and an explicit formula, and discussed various algebraic and differential properties. As a potential application, we calculated the Finsler metric of the most general spherically symmetric Finsler space and expressed it in terms of harmonic d-tensors.

Since d-tensors are abundant in spray and Finsler geometry, there are various possible applications of the harmonic d-tensors we presented here. Note that these applications are not restricted to three-dimensional manifolds. Spherical symmetry plays a role also in higher dimensions, for example in four dimensions with Lorentzian signature, where one of the dimensions carries the interpretation of time instead of space, and where a suitable decomposition into time and space can be performed. Note that this decomposition applies to d-tensors in the same way as to ordinary tensors on the manifold.

Various possible extensions of this work may be considered. The most interesting would be to extend our notion of harmonic objects also to connections on the tangent bundle instead of d-tensors, where two notions must be distinguished. The first notion is that of a non-linear connection on the tangent bundle, which can be expressed in terms of a tensor field of rank \((1,1)\) on \(TM\), and could therefore be addressed by decomposing tensors over \(TM\) into irreducible representations of the rotation group. An alternative definition, which is more closely related to the pullback bundle approach we used here, is to define a non-linear connection as a splitting of the exact sequence~\eqref{eq:exseq}. The other notion is that of a $N$-linear connection, which allows taking the covariant derivative of a d-tensor field, at whose components could be expressed in terms of the adapted basis shown in section~\ref{ssec:basis}.

Another possible extension of our work would be to consider other symmetry groups, such as the groups \(\mathrm{SO}(4)\), \(\mathrm{ISO}(3)\) and \(\mathrm{SO}(3,1)\) relevant in cosmology. For example, in the case of \(\mathrm{SO}(4)\) symmetry, one may use the coordinates introduced in~\cite{Hohmann:2015duq}, which have properties similar to the ones we have used here. Another possible line or research is to exploit the relation between Finsler and Cartan geometry discussed in~\cite{Hohmann:2013fca,Hohmann:2015pva} and to discuss Cartan geometries with particular symmetry groups.

\section*{Acknowledgments}
The author gratefully acknowledges the full support by the Estonian Ministry for Education and Science through the Institutional Research Support Project IUT02-27 and Startup Research Grant PUT790, as well as the European Regional Development Fund through the Center of Excellence TK133 ``The Dark Side of the Universe''.

\appendix


\end{document}